\documentstyle[12pt]{article}

\newcommand{\Od}{{\cal O}}

\newcommand{\tr}{\mbox{tr}}

\newcommand{\fpi}{f_\pi}
\newcommand{\fpite}{f_\pi^t}
\newcommand{\fpisp}{f_\pi^s}
\newcommand{\ft}{f(t)}

\newcommand{\fdot}{\dot f(t)}
\newcommand{\fddot}{\ddot f(t)}

\newcommand{\intc}{\int_C dt \int d^3 \vec{x}}

\newcommand{\vxt}{(\vec{x},t)}

\newcommand{\be}{\begin{equation}}
\newcommand{\ee}{\end{equation}}
\newcommand{\ba}{\begin{eqnarray}}
\newcommand{\ea}{\end{eqnarray}}

\newcommand{\NP}[1]{{\it Nucl.\ Phys.\ }{\bf #1}}
\newcommand{\ZP}[1]{{\em Z.\ Phys.\ }{\bf #1}}

\newcommand{\PL}[1]{{\em Phys.\ Lett.\ }{\bf #1}}

\newcommand{\AN}[1]{{\em Ann. Phys. } (N.Y.) {\bf #1}}

\newcommand{\PR}[1]{{\em Phys.\ Rev.\ }{\bf #1}}

\newcommand{\IJmp}[1]{{\em Int.\ J.\ Mod.\ Phys.\ }{\bf #1}}

\def\be{\begin{equation}}
\def\ee{\end{equation}}
\def\bea{\begin{eqnarray}}
\def\eea{\end{eqnarray}}

\begin{document}

\begin{center} 
{\large\bf NONEQUILIBRIUM 
CHIRAL PERTURBATION THEORY}\footnote{Proceedings of 
{\it Strong and Electroweak Matter 98}, Nordita 
 (Copenhaguen) 2-5 Dec. 1998}\\
\vskip 1.2cm
{\large 
A. G\'{o}mez Nicola$
$\footnote{E-mail:{\tt  gomez@eucmax.sim.ucm.es}}
}\\
\vskip 5pt 
Departamento de F\'{\i}sica 
Te\'orica,  Universidad Complutense, 28040, Madrid, Spain
\vskip 3pt
\end{center}

%
%


\begin{abstract}
We explore the extension of chiral
 perturbation theory to a meson gas out of thermal equilibrium.
 For that purpose, we let the pion decay constant be a time-dependent
 function  and work within the
 Schwinger-Keldysh contour technique. A useful connection with curved space-time
 QFT is established, which  allows to consistently renormalise the model.
 We discuss the applicability of our approach within  a heavy-ion collision
 environment
\end{abstract}

The  nonequilibrium dynamics of the chiral phase
 transition has attracted considerable interest during the last
 few years. One of the original motivations to analyse this regime
 was the suggestion  \cite{an89}
that the so called disoriented chiral condensates
 (DCC) could form during the plasma expansion after a relativistic
  heavy-ion collision (RHIC), giving rise to observable
 effects such as coherent pion emission \cite{rawi93}. Traditionally,
 this and other similar
 effects have been investigated using $O(N)$ models in which initial
 thermal equilibrium is assumed and
 nonequilibrium is parametrised   as the time dependence of the
 different lagrangian parameters
\cite{rawi93,bodeho95,cooper95,lamp96}. Such time dependence is indeed
 the only relevant nonequilibrium feature
when working for instance within Bjorken's initial conditions approach,
where observables such as
the order parameter  depend only on proper time in the central
 rapidity region \cite{bjo83}.

The use of effective models for QCD is imperative to
describe properly the dynamics of the plasma in this regime of
 temperature and energy density. The $O(N)$ models include
 explicitly the $\sigma$ meson and are valid  only for two
 light flavours ($N_f=2$). Besides, for strongly coupled systems,
 perturbation theory in these models is not well defined and one
 needs to use resummation methods such as large $N$.
 An alternative, not so well investigated, is to use
 the effective chiral lagrangian formalism, whose  main advantage
 is to provide  a consistent perturbation theory
in powers of $p/\Lambda_{\chi}$-- Chiral Perturbation Theory (ChPT)
 \cite{wegale}--
 where $p$ denotes generically
 any meson external momenta or field derivative,
 and $\Lambda_{\chi}\simeq$ 1 GeV.
 Besides, it is also valid for  $N_f=3$. So far, this formalism has been
 applied only in thermal equilibrium, to
 study  the low $T$ ($T=\Od(p)$) meson gas
 and the chiral phase transition  \cite{gele89,boka96}.

 We will review here recent work \cite{agnvgg99} on the extension of
 ChPT to a nonequilibrium situation. The key idea is to make use of the
 derivative expansion naturally incorporated in ChPT, in order to
 study the system not far from equilibrium. Of course, in a later stage
 one could apply  resummation methods such as large $N$ so as to extend
 the validity of the approach closer to the critical point, as it has
 already been done in equilibrium \cite{boka96}.

 Our starting point is the nonlinear sigma model (NLSM) where we
 let the pion decay constant--the only relevant parameter to the lowest
 order in derivatives--be time dependent. As commented above, such time
 dependence can
 be thought of as proper time evolution. We take the
  initial time $t=0$, which would correspond
   to a proper time $\tau_0\simeq$ 1 fm/c, a typical
 hadronisation time in a RHIC environment \cite{bjo83}.
  Thus, we will consider the NLSM action
\begin{equation}
S[U]=\intc \ \frac{f^2(t)}{4} \ \tr \ \partial_\mu U^\dagger\vxt
\partial^\mu U\vxt
\label{nlsmne}
\end{equation}

Here, $C$ is the Schwinger-Keldysh contour in time, containing an
 imaginary  leg of length $\beta_i=1/T_i$,  $T_i<T_c$ being the
 temperature for $t\leq 0$, where we assume that the system is in
 thermal equilibrium.
Thus, $f(t\leq 0)=\fpi\simeq$ 93 Mev
 to leading order. For $t>0$ the system departs from equilibrium.
 Note that, since we choose that departure to be instantaneous,
 $\ft$ cannot be analytical at $t=0$. This feature can give rise to
 discontinuities in the observables, and even
 extra singularities at $t=0$, as it has been noted in \cite{baac98}.
 Finally, in the above equation, $U(x)$ is the Goldstone boson field,
 which we can parametrise as customarily in terms of pions for $N_f=2$
plus kaons and eta for $N_f=3$ \cite{agnvgg99}. We shall restrict here
 to $N_f=2$.

The new ingredient we need to incorporate in the power counting in order
 to be consistent with ChPT is
\begin{equation}
\frac{\fdot}{f^2 (t)}\simeq \Od\left(\frac{p}{\Lambda_\chi}\right), \qquad
\frac{\fddot}{f^3(t)}, \frac{[\fdot]^2}{f^4(t)}
\simeq\Od \left(\frac{p^2}{\Lambda_\chi^2}\right),
\label{chipoco}
\end{equation}
and so on. Otherwise, we shall keep $\ft$ arbitrary. One can think
of $\ft$ as an external source,  to which we will find the
nonequilibrium
 response of the system. Another alternative, which we will not attempt
 here,  is  to treat $\ft$ as a field and solve for $\ft$ the
 hydrodynamic equations self-consistently.

Once we have defined our nonequilibrium power counting, we can apply
 ChPT to calculate the time evolution of the different observables.
 In doing so, we must pay special attention to renormalisation.
 In standard ChPT, one-loop UV divergences coming from the
 $\Od(p^2)$ lagrangian
 are canceled
 by tree level contributions coming from the $\Od(p^4)$ one, and so on for
 higher order contributions.
 Such fourth-order action is the most general one preserving all the
 symmetries. On the other hand, it is a well-known feature of nonequilibrium
 field theories that new infinities (time dependent in this case) can arise
 \cite{baac98,boleesi93}. Thus, our fourth-order lagrangian must contain
 necessarily new terms, to account for extra divergences. In order to find
 them, we will make use of a very fruitful analogy: the action
 (\ref{nlsmne}) is equivalent to formulate the
   NLSM on a curved space-time background
 corresponding to a spatially flat Robertson-Walker metric, with
 scale factor  $a(t)=f(t)/f(0^+)$ \cite{agnvgg99}.
 In this way, we can construct the
 $\Od(p^4)$ lagrangian in the following way: we just raise and
 lower indices with our RW metric in the standard (equilibirum)
 terms and add those
 terms coupling $U(x)$ with the scalar curvature $R(x)$ and the Ricci
 tensor $R_{\mu\nu} (x)$ to this order.
 The latter involve new low-energy constants that
 are fixed by analysing the energy-momentum tensor of QCD at low energies
 \cite{dole91}.

As a first application of our approach, we have analysed  the pion
 decay functions (PDF) to next to leading order, i.e, including loops
 with (\ref{nlsmne}) and tree level diagrams from the $\Od(p^4)$ lagrangian.
 In thermal equilibrium, one can define two pion decay constants, $\fpisp (T)$
 (spatial) and $\fpite (T)$ (temporal)
due to
 the loss of Lorenz covariance \cite{pity96}. The same happens
 out of equilibrium, where those constants turn into  time-dependent
 functions. After analysing the corresponding Feynman graphs
 the result can be written as \cite{agnvgg99}

\ba
\left[\fpisp (t)\right]^2&=&f^2 (t)\left[1+2f_2(t)-f_1(t)\right]-2i G_0(t)
\label{fpispnlo}\\
\left[\fpite (t)\right]^2&=&f^2 (t)\left[1+f_2(t)\right]-2i G_0(t)
\label{fpitenlo}
\ea
for $t>0$, where

\ba
f_1(t)&=&12\left[\left(2L_{11}+L_{12}\right)\frac{\fddot}{f^3(t)}
-L_{12}\frac{[\fdot]^2}{f^4 (t)}\right]
\nonumber\\
f_2(t)&=&4\left[\left(6L_{11}+L_{12}\right)\frac{\fddot}{f^3(t)}
+L_{12}\frac{[\fdot]^2}{f^4 (t)}\right]
\label{f12}
\ea

Here, $L_{11}$ and $L_{12}$ are the two new low-energy constants that need
 to be introduced to this order \cite{dole91}. While $L_{12}$ is already
 finite, $L_{11}$ needs to be renormalised in dimensional regularisation.
In (\ref{fpispnlo})-(\ref{fpitenlo}), $G_0 (t)$ is nothing but the
 equal-time pion two-point function  $G_0(t)=G_0(x,x)$ with
 $G_0(x,y)$ the solution of the differential equation
\begin{equation}
\left\{\Box_x + m^2(x^0)\right\} G_0 (x,y)=-\delta_C (x^0-y^0)
\delta^{(3)} (\vec{x}-\vec{y})
\label{loprop}
\end{equation}
with   KMS equilibrium  conditions $G^>_0
(\vec{x},t_i-i\beta_i;y)=G^<_0 (\vec{x},t_i;y)$, $t_i<0$ and
 $m^2(t)=-\fddot/\ft$ plays  the role of a time-dependent mass. It is
 important to remark that this mass is a consequence solely of the
 nonequilibrium behaviour and has nothing to do with a explicit chiral
 symmetry breaking pion mass term. In fact, our model
 is exactly
chiral invariant, since, for simplicity, we have chosen to work in the
 chiral limit (there are no pion mass terms in (\ref{nlsmne})).
  In the language of curved space-time QFT,
 $m^2(t)$ is the minimal coupling with the metric preserving chiral
 invariance. Note that $m^2(t)$ can be negative, thus allowing the
 existence of unstable long-wavelength modes, which play an essential
 role during the plasma evolution
\cite{rawi93,bodeho95,cooper95,lamp96}.

The results (\ref{fpispnlo})-(\ref{fpitenlo}) reproduce the
equilibrium $\fpi (T)$ when we switch off  the time derivatives of
$\ft$. We remark that $G_0(t)$ contains UV divergences,
 giving  rise to  time-dependent singularities,
 to be absorbed by $f_1(t)$
 and $f_2(t)$  in the renormalisation of $L_{11}$. An interesting
 consequence of our result is that  $\fpisp (t)\neq \fpite (t)$ to
 one-loop, unlike the equilibrium case \cite{pity96}. In addition,
 from (\ref{fpispnlo})-(\ref{fpitenlo}) and (\ref{f12}) we see that the
 difference $[\fpisp (t)]^2-[\fpite (t)]^2$ is finite, so that we can
 renormalise both at the same time, which is another consistency check.

Therefore,  given $\ft$, all one has to do to this order is to
 solve the leading order  propagator equation (\ref{loprop}). Of
 course, that is not possible in general, as it is well-known
  in the context of curved space-time QFT, where
   solutions can be analytically found  only for a very few
 choices of the scale factor, or equivalently, for $\ft$ \cite{bida82,sewe85}.
 Hence,
 in  order to estimate the nonequilibrium effects within our
 approach, we shall perform  a short-time approximation. We will
 restrict then to $0^+<t<t_{max}$ where  $0^+$ means  a small
 response time and $t_{max}\simeq 1/\fpi\simeq$ 2 fm/c. For that range,
 the expansion in $t$ is also a chiral expansion, so that we can truncate it
 to the relevant order. Thus,  solving  for the propagator by expanding
  $m^2(t)$ in power series, we get
\be
\left[\fpi^{s,t}(t)\right]^2=\left[f_R^{s,t}\right]^2
\left\{1-\frac{T_i^2}{T_c^2}
+2Ht
-\left[m^2\left(1-\frac{T_i^2}{T_c^2}\right)-H^2\right]t^2
+\Od(p^3/\Lambda_\chi^2)\right\}
\label{nlocoeff}
\ee
where $T_c=\sqrt{6}\fpi$,
$m^2=-\ddot f(0^+)/f(0^+)$ and $H=\dot f(0^+)/f(0^+)$. Notice that
 unstable modes appear for $m^2<0$. Here, the two constants $f_R^{s,t}$
 are already renormalised in terms of $L_{11}$ and $L_{12}$ and are fixed
 by the physical value of  $f_\pi^{s,t}(t=0^+)$. We remark that the
   $t=0$ discontinuities are a consequence of our choice of initial conditions.
 Nonetheless,  the effect of the $L's$ turns out to be relatively small since
  $L_{11}^r,L_{12}^r\simeq 10^{-3}$ \cite{dole91}.

Within our short-time approach, we can estimate the thermalisation time
 by $\fpi (t_f)=f$ (neglecting the effect of the $L's$) which is also
 the freezing time, since $\fpi$ reaches its zero temperature value.
 It is clear that by expanding in $t$ we cannot reproduce a thermalisation
 process
 where  $T_f\neq 0$. The maximum value
 $t_f\simeq t_{max}$ is reached for $H<0$ and $m^2<0$. In fact,
 unstable modes always tend to cool down the system. Our estimates
 are
  somewhat lower that  typical $O(N)$ calculations \cite{cooper95,lamp96}, which
   is not surprising, since in those works the
  $T_i\geq T_c$ and then gradients are too high to compare
 with our approach. For instance, in \cite{lamp96}, $T_i\simeq$ 200 MeV
 and $|H|\simeq$ 400 MeV. Thus, we expect our model  to be valid when
 some cooling has already taken place and the system enters the validity
 range of low-energy ChPT. Nevertheless, it is worth remarking  that
  large $N$ methods would allow us  to depart further from
 equilibrium. Other extensions and applications of our model to be analysed
 in the future include the long-time evolution -- by choosing suitable
 parametrisations for $\ft$--, the $N_f=3$ case and   the formation of DCC.
 The
inclusion of  quark masses  and gauge fields would allow   to
investigate, respectively,
 the quark condensate time dependence and photon production in the pion sector.

\section*{Acknowledgments}

I wish to thank the organisers of the ``Strong and Electroweak Matter 98''
 conference  for  a  really enjoyable atmosphere.  Financial
 support from CICYT, Spain,  project AEN97-1693, is  acknowledged.


\end{document}